\documentclass[aps,twocolumn,showpacs,floats,superscriptaddress]{revtex4}
\usepackage{graphicx}
\usepackage{dcolumn}
\usepackage{color}
\usepackage{bm}
\begin{document}


\author{Nikolay  Prokof'ev }
\author{Boris Svistunov}

\affiliation{Department of Physics, University of Massachusetts,
Amherst, MA 01003, USA} \affiliation{Russian Research Center
``Kurchatov Institute'', 123182 Moscow, Russia}

\title{Bold Diagrammatic Monte Carlo: When Sign Problem is Welcome}

\begin{abstract}
We introduce a Monte Carlo scheme for sampling bold-line
diagrammatic series specifying an unknown function in terms of
itself. The range of convergence of this bold(-line) diagrammatic
Monte Carlo (BMC) is significantly broader than that of a simple
iterative scheme for solving integral equations. With BMC technique,
a moderate ``sign problem" turns out to be an advantage in terms of
the convergence of the process. For an illustrative example, we
solve one-particle $s$-scattering problem. As an important
application, we obtain $T$-matrix for a fermipolaron (one spin-down
particle interacting with the spin-up fermionic sea).

\end{abstract}






\maketitle

Diagrammatic Monte Carlo (DiagMC) \cite{DiagMC} is a technique that
allows one to simulate quantities specified in terms of convergent
diagrammatic sums, i.e., sums of integrals with integrands
represented by a diagrammatic structure. Formally, it is a set of
generic prescriptions for organizing a systematic-error-free
Metropolis-type process that samples the series/sum without
explicitly taking the integrals over the internal variables in each
particular term. In addition to the natural requirement of
convergence, the diagrammatic sums should be either essentially
finite (have only a few leading terms) or positive definite;
otherwise the sign problem suppresses the efficiency of the numeric
procedure. One of the key tools in the analytical diagrammatic
techniques is the trick of bold lines \cite{bold_line} that allows
one to (partially or completely) sum the series even if it is
formally divergent. The bold-line trick looks also very attractive
for the sign-indefinite series since it can substantially reduce the
number of leading diagrams, and thus alleviate the sign problem.

In this Letter, we explore the possibility of employing the
bold-line trick in the DiagMC approach. We propose a scheme which
we call bold(-line) diagrammatic Monte Carlo (BMC). In essence,
BMC is a generalized iterative scheme  in which the iteration
protocol depends on the number of iteration steps, or,
equivalently, in which the next iteration is a function of not
only its immediate ancestor, but of the (properly weighted) whole
list of earlier iterations. The crucial difference between BMC and
a na\"{i}ve iteration protocol---when one simply uses DiagMC to
perform an integration for a given iteration step---is that the
convergence of BMC has essentially broader parameter range. We
present general arguments why this is the case and perform an
illustrative simulation for one-particle $s$-scattering problem.
Despite its formal simplicity, the problem contains all the
ingredients one can meet in a general case: the perturbative
series diverges if the scattering potential is strong enough,
and---in the case of a repulsive potential---the series is not
positive definite. The simplifications which we have here are
mainly quantitative rather than qualitative. The bold-line trick
reduces the infinite perturbative series to just two terms. In the
case of a strong attractive potential, two more terms appear in
the right-hand side to secure the convergence. Incidentally, with
these extra terms a sign problem arises for strong attractive
potential as well.

Before turning to the realistic simulation, we start with
discussing a generalized iterative procedure which is most close
to the actual BMC scheme.  To analyze the convergence issues, we
can confine ourselves with a linearized problem and write
\begin{equation}
|\, f\, \rangle \; =\;|\, b\, \rangle\, +\, {\bf A}\, |\, f\,
\rangle \; , \label{eq0}
\end{equation}
where $|\, f\, \rangle$ is some unknown vector, $|\, b\, \rangle$
is a known vector, and ${\bf A}$ is a linear operator. Expanding
all the vectors in terms of the operator ${\bf A}$ eigenvector
basis $\{|\, \phi_{\xi} \, \rangle \}$, we get
\begin{equation}
f_{\xi}\;  =\; b_{\xi}\, +\, a_{\xi}\, f_{\xi} \; , \label{eq1}
\end{equation}
where ${\bf A} |\, \phi_{\xi} \, \rangle  = a_{\xi}  |\,
\phi_{\xi} \, \rangle$, ~~$|\, b\,  \rangle = \sum_{\xi} b_{\xi}\,
|\, \phi_{\xi} \, \rangle$, ~~$|\, f\,  \rangle = \sum_{\xi}
f_{\xi}\, |\, \phi_{\xi} \, \rangle$. The  vector equation thus
decouples into a set of independent equations for each $\xi$. From
now on the label $\xi$ can be omitted. From the convergence point
of view, it is advantageous to work with iterative schemes that
involve---at least at the theoretical level---an averaging of the
iterations. Let $f_n$ be the $n$-th generator for the $(n+1)$-st
iteration
\begin{equation}
\tilde{f}_{(n+1)}\;  =\; b\, +\, a\, f_{n} \; . \label{iter}
\end{equation}
The quantity $f_n$ is supposed to be a function of all
$\tilde{f}_{j}$'s with $j\leq n$. As a characteristic example we
choose
\begin{equation}
f_n\; =\; { \sum_{j=1}^n\, j^{\alpha} \,\tilde{f}_j \over
\sum_{j=1}^n\, j^{\alpha}}
 \label{f_n}
\end{equation}
where $\alpha > -1$ is a fixed parameter of the scheme. We can
exclude $\tilde{f}$'s and explicitly relate $f_{(n+1)}$ to $f_{n}$
to see that for the deviation $\delta_n=f-f_{n}$ of the $n$-th
generator from the exact solution $f=b/(1-a)$ the following
recursive relation takes place
\begin{equation}
\delta_{(n+1)} \; =\; \delta_n\, +\,  {(1+\alpha)(a-1)\over n+1}\,
\delta_n \; . \label{delta_n}
\end{equation}
It implies the asymptotic behavior
\begin{equation}
\delta_n \; \propto \; {\rm e}^{(1+\alpha)(a-1)\ln
n}~~~~~~~~~~~(n\, \to\, \infty) \; . \label{asympt}
\end{equation}
Hence, the condition of convergence is
\begin{equation}
{\rm Re}\, a_{\xi} \, <\, 1 \; .  \label{converg}
\end{equation}
Here we restore the subscript $\xi$ to emphasize that condition
(\ref{converg}) has to be satisfied for all the eigenvalues of the
matrix ${\bf A}$. We see that the value of $\alpha$ does not
determine the fact of convergence, but does effect the asymptotic
rate of convergence---the larger is $\alpha$, the higher the rate.
It is important that the convergence does not depend on the
imaginary parts of $a_{\xi}$'s. Finally, {\it negative} real parts
of $a_{\xi}$ are {\it desirable} for convergence: the larger is
the absolute value of the negative real part, the better. [Note
that the plain iterative method ($f_n\equiv \tilde{f}_n$)
converges only when $|a_{\xi}|<1$.]

If condition (\ref{converg}) is not met, one can use an equation
equivalent to (\ref{eq0}), but with convergent iterative
procedure. For the $s$-scattering problem, the matrix ${\bf A}$ is
Hermitian and thus all its eigenvalues  are real. In this case,
rewriting Eq.~(\ref{eq0}) as
\begin{equation}
|\, f\, \rangle \; =\;|\, b\, \rangle\, +\, {\bf A}\, |\,
f\,\rangle \,+\, \lambda  {\bf A} \left( \, |\, f\, \rangle \, -\,
|\, b\, \rangle\, - \, {\bf A}\, |\, f\,\rangle \,  \right)
\label{eq3}
\end{equation}
and fine-tuning the value of the constant $\lambda$, one can
render the iterative process converging. Indeed, the new equation
has the same form as the original one, up to replacements $|\, b\,
\rangle\, \to\, |\, b\, \rangle \, -\, \lambda {\bf A} |\, b\,
\rangle$ and ${\bf A}\, \to \, (1+\lambda){\bf A}-\lambda {\bf
A}^2$. Correspondingly, condition (\ref{converg}) for the new
matrix will be met if the original eigenvalues satisfy the
inequality
\begin{equation}
(1+\lambda)a_{\xi}-\lambda a_{\xi}^2\, < \, 1 \; .
\label{converg1}
\end{equation}
As is easily checked, condition (\ref{converg1}) is met provided
$\lambda\, \in\,  (\lambda_1,\,  \lambda_2)$, where
\begin{equation}
\lambda_1^{-1}\, =\, \min_{a_{\xi}>1}\,  \{ a_{\xi} \}\, ,
~~~~~~~~\lambda_2^{-1}\, =\,\max_{a_{\xi}\in (0,1)}\,  \{ a_{\xi}
\}\; . \label{lambda12}
\end{equation}
Hence, if the eigenvalues are real and separated from unity by a
finite gap, there exists a value of $\lambda$ at which convergence
is guaranteed. Incidentally, the problem (\ref{eq0}) can always be
re-formulated in such a way that the new matrix is Hermitian:
\begin{equation}
|\, f\, \rangle \; =\;(1-{\bf A}^{\dagger})\, |\, b\, \rangle\,
+\, ({\bf A}+{\bf A}^{\dagger}-{\bf A}^{\dagger}{\bf A})\, |\, f\,
\rangle \; . \label{eq4}
\end{equation}

The $s$-scattering problem can be formulated (see, e.g.,
Ref.~\cite{LL}; for simplicity, we work with a spherically symmetric
potential) as Eq.~(\ref{eq0}) with $|\, f\, \rangle \equiv f(q)$
being the zero-energy scattering wave function in the momentum
representation. In this case $|\, b\, \rangle \equiv -u(q)$, where
$u(q)=U(q)/2\pi$, and $U(q)$ is the Fourier transform of the
scattering potential; the Plank's constant and particle mass are set
equal to unity. The operator ${\bf A}$ here is the integral operator
\begin{equation}
{\bf A}f\; =\; -\, {1\over \pi}\int_{-1}^{1}d\chi \int_0^{\infty}
u\left( |{\bf q}-{\bf q}_1| \right) \, f(q_1) \, dq_1 \; ,
\label{A_f}
\end{equation}
where $|{\bf q}-{\bf q}_1|\equiv \sqrt{q^2+q_1^2-2qq_1\chi}$;
Eq.~(\ref{eq3}) then reads
\begin{equation}
f\; =\; -u+\lambda {\bf A}u+(1+\lambda){\bf A}f-\lambda{\bf A}^2 f
\; . \label{eq33}
\end{equation}
The potential we use in simulations is a flat spherical well/bump
defined as $U({\bf r})=U_0$ at $r<1$ and zero otherwise. For this
potential,
\begin{equation}
u(q)\; =\; {2U_0\over q^3}\, ( \sin q-q\cos q )  \; .
\label{potential}
\end{equation}

\noindent{\it Monte Carlo procedure.} Here we discuss how generic
DiagMC rules are used to calculate $f(q)$ self-consistently. For
brevity, let us index the four terms in the right-rand-side of
Eq.~(\ref{eq33}) as terms ${\cal A}$, ${\cal B}$, ${\cal C}$, and
${\cal D}$. Correspondingly, the ``configuration space" of the
problem is defined by the term index and all continuous variables
associated with it. The goal of the Monte Carlo procedure is to
perform stochastic summation over this configuration space. The
contribution of each state to the answer is characterized by the
weight with the sign which in our case is given by the product of
$u$- and $f$-functions; for example, the weight and sign of the
$({\cal B}, q,q',\chi)$ state is determined by the modulus and
sign of $u(|{\bm q}-{\bm q}'|)f(q')$.

The standard Metropolis-type protocol consists of updates which
change the current configuration state followed by measurements
which evaluate contributions of the current state to the answer. The
updating scheme described below generates states with probabilities
proportional to their weight. In this case, the Monte Carlo
estimator for $f(q)$ is given by the state sign. The statistics of
$\pm 1$ contributions is accumulated into the $f(q)$-histogram with
bins covering the positive-$q$ axis. Apart from representing the
final result of the simulation, the histogram is used
self-consistently to generate random variables from the probability
density $|f(q)|$ and to determine the sign of ${\cal B}$ and ${\cal
D}$ states.

A straightforward accumulation of data into the histogram
corresponds to $\alpha=1$ in Eq.~(\ref{f_n}). However, large
values of $\alpha$ result in faster convergence, see
Eq.~(\ref{asympt}). Numerically, the limit of $\alpha \to \infty$
is implemented by simply erasing ``old" histogram data.

The simplest updating scheme contains three pairs of complementary
updates $[{\cal A} \to {\cal B},{\cal B} \to {\cal A}]$, $[{\cal
A} \to {\cal C},{\cal C} \to {\cal A}]$, $[{\cal C} \to {\cal
D},{\cal D} \to {\cal C}]$ which change the term index, and one
self-complementary update ${\cal A}\leftrightarrow {\cal A} $
changing the variable $q$.

${\cal A}\leftrightarrow {\cal A} $ {\it update}. A  new value for
the variable $q$ in state ${\cal A}$ is generating from the
normalized probability density
\begin{equation}
p(q)=|u(q)|/I_u\;, \;\;\;\;\; I_u=\int_0^{\infty} |u(q)| \, dq \;,
\label{pu}
\end{equation}
The acceptance ratio for the ${\cal A}\leftrightarrow {\cal A} $
update is unity.

${\cal A}\to {\cal B}$ {\it update}. First, we select the value of
$\chi$ from the uniform probability density on the $[-1,1]$
interval. Next, we select the value $q'$ from the histogram based
probability distribution $|f(q')|$. The acceptance ratio for this
update is
\begin{equation}
R_{{\cal A}\to {\cal B}} = {2|1+\lambda|\, I_f\over \pi p_{\cal
AB} }\, \left| { u({\bm q}-{\bm q}')\over u(q)}\right|\; ,
\label{AB}
\end{equation}
where $p_{\cal AB}$ is the probability to apply the ${\cal A}\to
{\cal B}$ update while in state ${\cal A}$ (we do not mention the
probability of applying an update to the current configuration if
it is unity; in this scheme $p_{\cal BA}=1$).
\begin{equation}
I_f\; =\;  \int_0^{\infty} | f(q)| \, dq  \label{I_f}
\end{equation}
is the normalization integral proportional to the sum of absolute
values of all histogram contributions (its proper normalization is
discussed below).

${\cal B}\to {\cal A}$ {\it update}. Here we propose to change the
term index back to ${\cal A}$; the acceptance ratio for this move
is simply the inverse of $R_{{\cal A}\to {\cal B}}$.

{\it ${\cal A}\to {\cal C}$ and ${\cal C}\to {\cal A}$ updates}.
Formally, this pair of updates is identical in implementation to
the previous one with the only difference that the value of $q'$
in the ${\cal A}\to {\cal C}$ move is generated from the
probability density $|u(q')|$. Correspondingly, the acceptance
ratio is based on the $I_u$ integral:
\begin{equation}
R_{{\cal A}\to {\cal C}} = {1 \over R_{{\cal C}\to {\cal A}}} =
{2|\lambda |\,  I_u \, p_{\cal CA}\over \pi \, p_{\cal AC} }\,
\left| { u({\bm q}-{\bm q}')\over u(q)}\right|\; , \label{AC}
\end{equation}

{\it ${\cal C}\to {\cal D}$ and ${\cal D}\to {\cal C}$ updates}
are an exact copy of the ${\cal A}\to {\cal B}$ and ${\cal B}\to
{\cal A}$ pair in terms of how new variables are proposed and
removed. The acceptance ratio is
\begin{equation}
R_{{\cal C}\to {\cal D}} = {1 \over R_{{\cal D}\to {\cal C}}} =
{2I_f\over \pi \, p_{\cal CD} }\, \left| { u({\bm q}'-{\bm
q}'')\over u(q')}\right|\; , \label{CD}
\end{equation}

The above set of updates is ergodic, i.e. it samples the entire
configuration space. In the practical implementation of the
algorithm we used $p_{\cal AA}=0.2$, $p_{\cal AB}=p_{\cal
AC}=0.4$, and $p_{\cal CA}=p_{\cal CD}=0.5$. To complete the
description we have to explain how to find the normalization
integral $I_f$ in Eq.~(\ref{I}). Let $Z_{\rm MC}$ be the total
number of Monte Carlo states in the simulation and $Z_A$ be the
number of ${\cal A}$-states. In the statistical limit,
\begin{equation}
{Z_{\cal A} \over Z_{\rm MC} }= { I_u \over I}\; , \label{Za/Zmc}
\end{equation}
\begin{eqnarray}
&&I=I_f+ {|1+\lambda| \over \pi} \int  |u({\bm q}-{\bm q}') f(q')|
d\chi dq dq'  \nonumber
\\
&+& {|\lambda| \over \pi} \int  |u({\bm q}-{\bm q}') u(q')| d\chi dq dq'  \label{I} \\
&+& {|\lambda| \over \pi^2} \int  |u({\bm q}-{\bm q}') u({\bm
q}'-{\bm q}'') f(q'')| d\chi d\chi' dq \,dq' dq'' \;.\nonumber
\end{eqnarray}
is the auxiliary ``global partition function" which drops out from
all final answers. If $H_s$ is the sum of all contributions to
the $s$-th bin of the histogram then in the statistical limit,
\begin{equation}
{H_s \over Z_{\rm MC} } = I^{-1} \int_{q\in {\rm bin}_s}f (q)\, dq
\; . \label{H_s}
\end{equation}
If we now write the normalization integral as a sum over the
histogram (in the limit of infinitesimally small bin size the
relation is exact)
\begin{equation}
I_f = \sum_s \left| \int_{q\in {\rm bin}_s}f_0(q)\, dq \right| =
(Z/Z_{\rm MC})\,  \sum_s\, |H_s| \; , \label{I_fa}
\end{equation}
and use Eq.~(\ref{Za/Zmc}) to eliminate $I/Z_{\rm MC}$ we finally
arrive at
\begin{equation}
I_f = { \sum_s |H_s| \over Z_{\cal A} }\,I_u \; . \label{I_est}
\end{equation}
Similarly, the physical result for the scattering wave function is
given by
\begin{equation}
f(q_s) = {H_s \over Z_{\cal A} }\,I_u \; . \label{f_est}
\end{equation}

The s-wave scattering length can be obtained in two ways: from the
$q\to 0$ limit, $a=-f(q=0)$, and as a histogram sum (the last
procedure gives better accuracy since it is based on all Monte
Carlo data and thus is not susceptible to the noise in a
particular bin)
\begin{equation}
a  = u(0) + {2\over \pi} \int_0^{\infty} \!\!\!\!  u(q) f(q) dq\,
\to\, u(0) + {2I_u \over \pi Z_{\cal A} } \sum_s  u(q_s) H_s \; .
\label{a2}
\end{equation}

\begin{figure}[tbp]
\centerline{\includegraphics [bb=0 -20 600 870, angle=-90,
width=0.95\columnwidth  ]{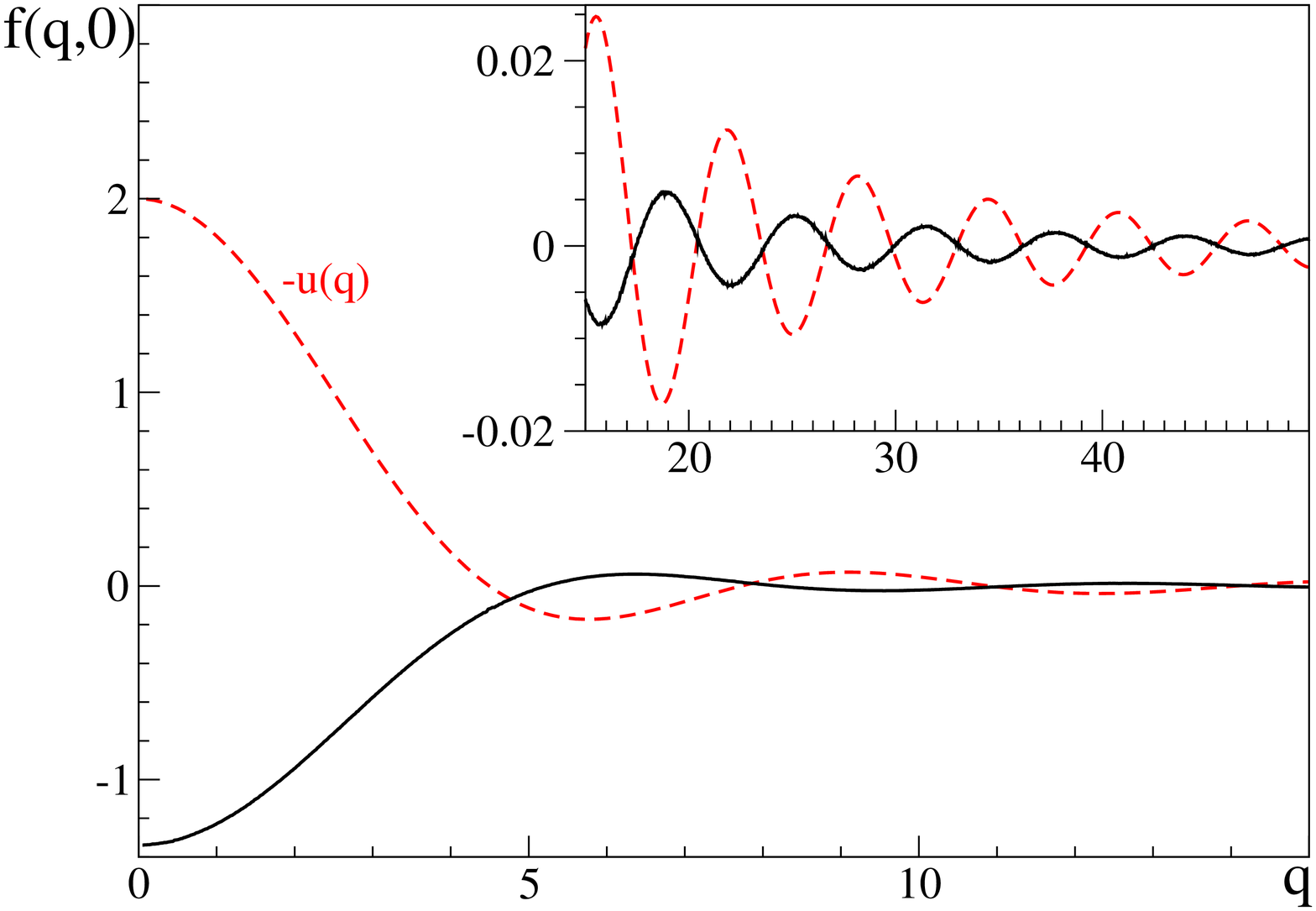}} \caption{(Color online).
Scattering wave function at zero energy (solid line) and
scattering potential (dashed line) for the attractive potential
well with one bound state ($U_0=-3$).} \label{fig1}
\end{figure}

We have tested our BMC scheme against the analytical answer for
the scattering length in different regimes which included strong
repulsive and attractive potentials outside of the convergence
limits for the standard iterative scheme. For example, one can
easily get results for $a$ with four-digit (or higher) accuracy
for the repulsive potential $U_0 = 10$; a straightforward
summation of the series expansion for large positive values of
$U_0$ would be impossible because of the divergence and/or the
sign problem. Series divergence will also prevent one from going
across the resonance and getting results for potentials with bound
states. In Fig.~\ref{fig1} we present data for the scattering wave
function obtained for $U_0=-3$, i.e. for the potential well with
the bound state. In this simulation $a$ was obtained with the
4-digit accuracy. For negative values of $U_0 < -10$ we found that
good initial conditions, e.g. results of the previous run for
smaller $|U_0|$, are important for convergence which was very slow
and required that $\lambda \approx 1$. In view of
Eq.~(\ref{lambda12}), we conclude that in this parameter range we
face the problem of having matrix {\bf A} eigenvalues being too
close to unity.

\begin{figure}[t]
\vspace*{-0cm}
\includegraphics[bb=65 380 535 430, width=\columnwidth]{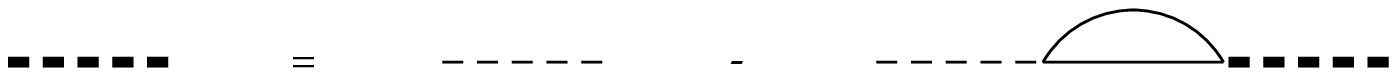}
 \caption{The diagrammatic equation for the $T$-matrix (heavy dashed line) in
 terms of the vacuum $T$-matrix (light dashed line), spin-down vacuum propagator
 (straight solid line), and truncated (to the momenta {\it less} than  Fermi momentum)
 spin-up propagator (solid arc). } \label{fig2}
\end{figure}

{\it Fermipolaron $T$-matrix}. The fermipolaron is a spin-down
particle interacting with the sea of spin-up fermions. Of special
interest is the case when the spin-up sea is an ideal Fermi gas
while the interaction between spin-up and spin-down particles is
short-ranged but resonantly strong. In this regime---relevant to the
notorious problem of BCS-BEC crossover in the limit of extreme
population imbalance between the two fermionic components
\cite{Lobo}---there is, in particular, a critical point in the
interaction strength when the groundstate of the polaron becomes a
bound spin-zero state (molecule). The fermipolaron problem allows an
unbiased numeric solution by DiagMC. The relevant diagrams are
constructed out of the three types of propagators (in the
imaginary-time--momentum representation): (i) spin-up Green's
functions, (ii) spin-down vacuum propagators, and (iii) the
$T$-matrix of the pair interaction between spin-up and spin-down
particles \cite{fermipolaron}. While simple analytic expressions for
type (i) and (ii) propagators are available, the T-matrix has to be
tabulated numerically; this tabulation represents the performance
bottleneck for the whole scheme. Noting that the equation defining
the $T$-matrix through itself (see Fig.~\ref{fig2}) is analogous to
that for the scattering amplitude, one can directly apply the
above-described BMC procedure for obtaining the $T$-matrix. We have
successfully done that, which ultimately allowed us to solve the
fermipolaron problem \cite{fermipolaron}.

{\it Conclusions and outlook}. We have found a numeric counterpart
to the bold-line trick of diagrammatic technique. The resulting
scheme simulates unknown functions specified by diagrammatic
series in terms of themselves. We illustrated our approach by
solving the $s$-scattering problem in strong repulsive and
attractive potentials. We introduced tools to secure convergence
of the process. With these tools we were able to solve the
$s$-scattering problem even in an attractive potential with a
bound state---an essentially non-groundstate problem.

The standard many-body diagrammatic technique deals with three
functions that are expressed through each other: Green's function,
self-energy, and the four-point vertex. The generalization of the
scheme to this case is theoretically straightforward, since formally
one can think of all these functions as different components of the
vector $| f \rangle $. The three practical questions that are
immediately seen---in the order of their importance---are: (i)
regularization of bold-line series, (ii) convergence of the scheme,
(iii) optimal data structure. Formally, the convergence of the
bold-line series may depend on the summation order and in certain
cases be achievable at the expense of controllable systematic error.
One may keep the expansion-order, $m$, of irreducible diagrams
finite and extrapolate results to the $m\to \infty$ limit, work with
a finite-size system and extrapolate to the thermodynamic limit,
introduce constraints on continuous variables not allowing them to
be either very small or very large and apply renormalization
techniques (ultra-violet divergences would be a typical example),
etc. The convergence of the scheme may be achieved by the tools
described in this paper. If the initial approximation to unknown
functions is close enough to the exact answer---which will be the
case if one starts with an almost ideal system and moves to a
strongly interacting regimes by small steps in the interaction
constant, then one may rely on linearization for constructing the
correcting part of the right-hand side operator, using prescriptions
of Eq.~(\ref{eq3}). If unknown functions depend on many continuous
variables, histograms may well be not the best method. Instead one
may use a variable-step meshes and, correspondingly, re-weighing
techniques for collecting statistics. Another option is to
approximate unknown functions with analytic expressions and
permanently optimize their parameters to best fit coarse-grained
histogram sampling coming from the BMC process.

The work was supported by the National Science Foundation under
Grants PHY-0426881 and  PHY-0653183. NP acknowledges support from 
PITP, Vancouver.

\end{document}